\documentclass[pre,twocolumn,floatfix,superscriptaddress,longbibliography]{revtex4-1}
\pdfoutput=1

\usepackage[utf8]{inputenc}

\usepackage{amsmath,amssymb,amsfonts}
\usepackage[colorlinks=true,citecolor=blue,linkcolor=magenta]{hyperref}

\usepackage{subcaption}
\usepackage[justification=centerlast,format=plain]{caption}

\usepackage[capitalize]{cleveref}

\usepackage{xcolor}

\usepackage{graphicx}
\usepackage{array}
\newcommand{\ket}[1]{\ensuremath{\left\vert #1 \right\rangle}}

\newcommand{\expectationvalue}[2]{\ensuremath{\left\langle #1 \middle\vert #2 \middle\vert #1 \right\rangle}}

\newcommand{\heff}{\ensuremath{h^{\text{eff}}}}
\newcommand{\hin}{\ensuremath{h^{\text{in}}}}
\newcommand{\thetain}{\ensuremath{\theta}}



\begin{document}
\title{Single-Qubit Cross Platform Comparison of Quantum Computing Hardware}

\author{Adrien Suau}
\affiliation{Theoretical Division, Los Alamos National Laboratory, Los Alamos, NM 87545, USA}
\affiliation{CERFACS, 42 Avenue Gaspard Coriolis, 31057 Toulouse, France}
\affiliation{LIRMM, University of Montpellier, 161 rue Ada, 34095 Montpellier, France}

\author{Jon Nelson}
\affiliation{Advanced Network Science Initiative, Los Alamos National Laboratory, Los Alamos, NM 87545, USA}

\author{Marc Vuffray}
\affiliation{Theoretical Division, Los Alamos National Laboratory, Los Alamos, NM 87545, USA}

\author{Andrey Y. Lokhov}
\affiliation{Theoretical Division, Los Alamos National Laboratory, Los Alamos, NM 87545, USA}
\author{Lukasz Cincio}
\affiliation{Theoretical Division, Los Alamos National Laboratory, Los Alamos, NM 87545, USA}
\author{Carleton Coffrin}
\affiliation{Advanced Network Science Initiative, Los Alamos National Laboratory, Los Alamos, NM 87545, USA}


\begin{abstract}
As a variety of quantum computing models and platforms become available, methods for assessing and comparing the performance of these devices are of increasing interest and importance.
Despite being built of the same fundamental computational unit, radically different approaches have emerged for characterizing the performance of qubits in gate-based and quantum annealing computers, limiting and complicating consistent cross-platform comparisons.
To fill this gap, this work proposes a single-qubit protocol (Q-RBPN) for measuring some basic performance characteristics of individual qubits in both models of quantum computation.
The proposed protocol scales to large quantum computers with thousands of qubits and provides insights into the distribution of qubit properties within a particular hardware device and across families of devices.
The efficacy of the Q-RBPN protocol is demonstrated through the analysis of more than 300 gate-based qubits spanning eighteen machines and 2000 annealing-based qubits from one machine, revealing some unexpected differences in qubit performance.
Overall, the proposed Q-RBPN protocol provides a new platform-agnostic tool for assessing the performance of a wide range of emerging quantum computing devices.
\end{abstract}


\maketitle

\section{Introduction}

In the current era of Noisy Intermediate-Scale Quantum (NISQ) devices \cite{Preskill2018quantumcomputingin}, a wide variety of technologies are being developed that leverage quantum mechanics to conduct, and hopefully accelerate, computations \cite{Ladd2010,Clarke2008,Johnson2011,Debnath2016}.
Given the noisy nature of these emerging technologies, measuring and tracking the fidelity of quantum hardware platforms is essential to understanding the limitations of these devices and quantifying progress as these platforms continue to improve.
Measuring the performance of gate-based quantum computers (QC) has been studied extensively through the topics of quantum characterization, verification, and validation (QCVV) \cite{Eisert2020}.
The scope of QCVV is broad and ranges from testing individual quantum operations (e.g., error rates of one- and two-qubit gates \cite{Wright2019}), verifying small circuits (e.g., Randomized Benchmarking \cite{PhysRevLett.106.180504,PhysRevA.77.012307}, Gate Set Tomography \cite{2009.07301}), to full system-level protocols (e.g., quantum volume estimation \cite{PhysRevA.100.032328}, random quantum circuits \cite{Boixo2018}).
Over the years these QCVV tools have become an invaluable foundation for benchmarking and measuring progress of quantum processors \cite{ibm_qv}, culminating with a quantum supremacy demonstration in 2019 \cite{Arute2019}.

Interestingly, this large body of QCVV work cannot usually be applied to the assessment of quantum annealing (QA) computers, such as the quantum devices developed by D-Wave Systems \cite{Johnson2011,dwave_docs}.
The fundamental challenge in conducting characterization, verification, and validation of quantum annealing devices (QAVV) is that available hardware platforms only allow measuring the state of the system in a fixed basis (the so-called computational $z$-basis) and at the completion of a specified annealing protocol.
Consequently, the QA user can only observe a fairly limited projection of the quantum state that occurs during the hardware's computation.
Despite these challenges, recent work has proposed the QASA protocol for performing single-qubit fidelity assessment of QA hardware platforms \cite{9465651}.

The notable differences in the computational models for QC and QA present significant challenges for conducting consistent comparisons of these hardware platforms.
Such efforts have focused almost exclusively on system-level benchmarks for applications like optimization and sampling featuring with 10s to 100s of qubits \cite{dw_qaoa_report,Willsch2020,10.1145/3425607,2107.06468}.
Conducting consistent comparisons of this type are notoriously difficult as these applications require the execution of complex quantum programs and often feature a large number of implementation parameters that impact performance measures, leaving significant room for debate around the fairness of these comparisons.
A key benefit in conducting QCVV/QAVV on small quantum systems (e.g., 1-3 qubits) is a reduction in protocol complexity that can reduce debate and increase confidence on the benchmark's results.

The core observation of this work is that the single-qubit QASA protocol, that was originally developed for tracking the performance of QA qubits \cite{9465651}, can be adapted for execution on the more general QC hardware platforms.
To that end, this work proposes the Qubit Response, Bias, Positive saturation, Negative saturation (Q-RBPN) protocol to measure four basic properties of the qubits in quantum hardware platforms.
%
%
In the broader context of QCVV, the Q-RBPN protocol is limited in the results it yields: however, its principal advantage is in providing a comparison of qubit performance that is agnostic to the underlying computational model.
To the best of our knowledge, this work provides the first side-by-side comparison of single-qubit performance across the QC (IBM-Q) and QA (D-Wave Systems) hardware platforms.

This work begins with a theoretical discussion of the computational tasks that underpins the Q-RBPN protocol and how it can be executed in different computing platforms in \cref{sec:sq_models}.
This explanation is followed by an analysis of more than 300 QC qubits and 2000 QA qubits in \cref{sec:full_chip}, with some discussion of the disintct features observed in different platforms.
\Cref{sec:conclusion} concludes the paper with a discussion of the value of universal qubit performance metrics and the broader implications of the results produced by the Q-RBPN protocol in this work.

\section{Single-Qubit Programming Models}
\label{sec:sq_models}

At first glance, the computational task used in this work is peculiar.
It is inspired by sampling from Gibbs distributions, a task that both QC and QA devices were not intended to solve.
However, as this work demonstrates, this unconventional task has the advantage of being realizable in both computational models in practice.
In the current NISQ era, we observe that this sampling task provides a useful measure of the consistency of qubit behaviors and output realizations in both QC and QA hardware platforms.

At the most abstract level, this work leverages a single input parameter (i.e., $\hin{}$) to control the relative amount of up/down measurements (i.e., $\sigma \in \{-1,+1\}$) that are observed in the output of the quantum device.
The hardware's performance is determined by how consistently it responds to different values of $\hin{}$ within a range of $-1.0$ to $1.0$.
Given the restrictions imposed by current QA hardware platforms this work only considers measuring spin projections in a fixed computational basis (denoted as the $z$-basis).

Drawing inspiration from \cite{9465651}, this work will focus on fitting Gibbs distributions to the outcomes of a single-qubit quantum programs, i.e.,
\begin{align}
    P(\sigma) &\propto \exp \left( \beta \hin{} \sigma \right). , \label{eq:gibbs_prob}
\end{align}
where $\sigma$ is an observed spin state, $\hin{}$ is an input parameter, and $\beta$ is a parameter corresponding to a thermal equilibrium at an effective temperature $1/\beta$. 
In QA, the $\beta$ parameter is determined by the details of the annealing protocol \cite{9465651}, whereas in QC it can be selected as part of the protocol.
The probability distribution over any single binary variable $\sigma \in \{-1,+1\}$ can be fully characterized by a single parameter $h^{\textrm{eff}}$, coined the effective field, in the following manner:
\begin{align}
\mathbb{P}\left(\sigma =  \pm1 \right) = \frac{\exp{\left( h^{\textrm{eff}}\sigma\right)}}{2\cosh{h^{\textrm{eff}}}}. \label{eq:effective_outcome_prob}
\end{align}
The value of $h^{\textrm{eff}}$ depends on the experiment's input parameters and is, in particular, a function of the input field $\hin{}$.
In the case of a classical magnet placed into a persistent external magnetic field $h$ in a thermal equilibrium at temperature $1/\beta$, one will observe a linear relationship between the output and input fields of the form $h^{\textrm{eff}} = \beta \hin{}$.
This linear mapping is called a \emph{classical} Gibbs distribution for a single spin; however, as noted in \cite{9465651}, the noisy nature of quantum hardware yields deviations from this linear mapping revealing imperfections and limitations of the hardware.
The remainder of this section details how the QC and QA platforms can be programmed to conduct single-qubit Gibbs sampling and how those observations are converted into $h^{\textrm{eff}}$ values.

\subsection{Theoretical Models}
\label{sec:theoretical-models}

\subsubsection{Gate-based Quantum Computing}
\label{sec:theoretical-model-gate-computing}

Our goal is to develop a quantum circuit that will include a free input parameter \hin{} and whose output will follow the Gibbs distribution from \cref{eq:gibbs_prob}. To this end, we transform a single qubit from its initialized state \ket{0} to a general state on the Bloch sphere,
\begin{equation}
\label{eq:bloch_state}
  \ket{\psi} = \cos{\left(\frac{\theta}{2}\right)} \ket{0} + e^{i\phi} \sin{\left(\frac{\theta}{2}\right)} \ket{1},
\end{equation}
and perform a measurement along the computational basis. The mapping is realized using the $1$-qubit gates $R_y(\theta) = e^{-i\theta Y}$ and $R_z(\phi) = e^{-i\phi Z}$, where $X,Y,Z$ denote the Pauli matrices. In order to obtain the linear relationship between the output \heff{} and input \hin{} fields, we prescribe the following mapping between the input parameter \hin{} and the angle $\theta$, 
\begin{equation}
  \label{eq:theta-hin}
  \thetain{} = \cos^{-1}\left( \tanh \left( \beta \hin{} \right) \right).
\end{equation}
This value of $\thetain{}$ depends on $\beta$ and \hin{}, and links the observed output field \heff{} and the input field \hin{} with the ideal linear relationship $\heff{} = \beta \hin{}$. 
This property means that, in an error-less QC model, one can compute an output field that is equal to $\beta \hin{}$ from the data collected on QC device.

\begin{figure}[t]
    \includegraphics{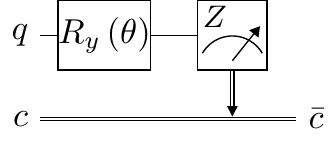}
    \caption{Generic $1$-qubit quantum program used by the Q-RBPN protocol on QC hardware. The rotation angle $\theta$ is related to an $h^{\textrm{in}}$ value according to \cref{eq:theta-hin}. 
}
    \label{fig:one-qubit-qc-theta}
\end{figure}

To see that the aforementioned linear relationship is indeed the theoretically expected one, observe that the expected value of a spin prepared in the state $\ket{\psi}$ from \cref{eq:bloch_state} and measured along the computational Z basis is,
\begin{align}
  \expectationvalue{\psi}{Z} 
  \label{eq:expected-value-ry}
                             &= \cos\left( \theta \right).
\end{align}
Substituting the expression of $\thetain{}$ from \cref{eq:theta-hin} in \cref{eq:expected-value-ry}, gives
\begin{equation}
  \label{eq:expected-value-hin}
  \expectationvalue{\psi}{Z} = \tanh\left( \beta \hin{} \right).
\end{equation}
Now, notice that the expectation of the measured output distribution from \cref{eq:effective_outcome_prob} is,
\begin{equation}
    \mathop{{}\mathbb{E}}\left[ \sigma \right] = \tanh\left( \heff{} \right).
\end{equation}
Therefore, we see that if the single spin system works ideally, one will obtain the following linear relationship between \hin{} and \heff{},
\begin{equation}
    \heff = \beta \hin{}.
\end{equation}
This linear relationship is, in theory, invariant with respect of the value of $\phi$. In \cref{sec:alternative-quantum-gate-program}, we show that, despite potential qubit imperfections, this invariance remained true in practice as well.
Based on this observation, the results of this work use the special case where $\phi=0$, which only requires $R_y$ rotations.
The quantum circuit that is ultimately utilized by the Q-RBPN protocol on QC hardware is depicted in \cref{fig:one-qubit-qc-theta}.

\subsubsection{Quantum Annealing}
\label{sec:theoretical-model-qa}

In contrast to QC, which can encode arbitrary quantum Hamiltonians, current quantum annealing platforms specialize in modeling the Hamiltonian of the transverse field Ising model, i.e.,
\begin{equation}
H_{\text{T-Ising}} =  \Gamma \sum_{i} X_i + \sum_{i,j} J_{ij} Z_i Z_j + \sum_{i}  h_{i} Z_i,
\end{equation}
where $h, J$ encode the local fields and interaction strengths of a classical Ising model and $\Gamma$ controls the strength of a global transverse field on that model.
This specialized Hamiltonian is of practical interest as the Ising model component can readily encode challenging computational problems arising in the study of magnetic materials, machine learning, and optimization \cite{hopfield1982neural,panjwani1995markov,lokhov2018optimal,Kochenberger2014} while the transverse field component yields complex entanglement structures that can be challenging to simulate \cite{quant-ph-0001106,10.5555/2011772.2011773}.

The quantum annealing protocol strives to find the low-energy assignments to a user-specified Ising problem by conducting an interpolation process of $H_{\text{T-Ising}} $ as follows:
\begin{multline}
    H_{\text{T-Ising}}(s) = \\ A(s) \sum_{i} X_i + B(s) \left( \sum_{i,j} J_{ij} Z_i Z_j + \sum_{i}  h_{i} Z_i \right). \label{eq:qah}
\end{multline}
The interpolation process starts with $s = 0$ and ends with $s = 1$. The two interpolation functions $A(s)$ and $B(s)$ are designed such that $A(0) \gg B(0)$ and $A(1) \ll B(1)$, that is, starting with a Hamiltonian dominated by the transverse field and slowly transitioning to a Hamiltonian dominated by an Ising model.
The outcome of this quantum annealing process is specified by a binary variable $\sigma_i$ that takes a value $+1$ or $-1$ and corresponds to the observation of the spin projection in the computational basis $Z$. 

In an idealized setting, and when the annealing interpolation transition process is sufficiently slow, the quantum annealing is referred to as adiabatic quantum computation.
The adiabatic theorem states that if the interpolation is sufficiently slow and the quantum system is isolated, the proposed QA protocol will always find the ground state of the $H_{\text{Ising}}$ problem \cite{quant-ph-0001106,PhysRevE.58.5355}.
However, existing QA hardware platforms are open-quantum systems \cite{PhysRevA.91.062320,Boixo2016,Smirnov_2018,dwave_docs} yielding outputs that are similar to thermal equilibrium distributions at an inverse effective temperature of $\beta \approx 10$ \cite{9465651,2012.08827}.

Focusing on the single-qubit context considered by this work, the QA Hamiltonian \eqref{eq:qah} reduces to,
\begin{equation}
    H_{\text{T-Ising}}(s) = A(s) \; X + B(s) \; \hin{} Z.
\end{equation}
Notice that the expectation of the measured output of this QA program can be used to compute \heff{} similarly to the QC case.
Despite the simplicity of this model, it has been observed that the imperfections of real-world QA platforms make it a useful tool for assessing the performance of individual qubits in practice \cite{9465651}.

\subsection{Illustration on Typical Qubits}

\begin{figure}[t]
    \centering
    \begin{subfigure}[t]{0.49\textwidth}
        \centering
        \includegraphics[width=\textwidth]{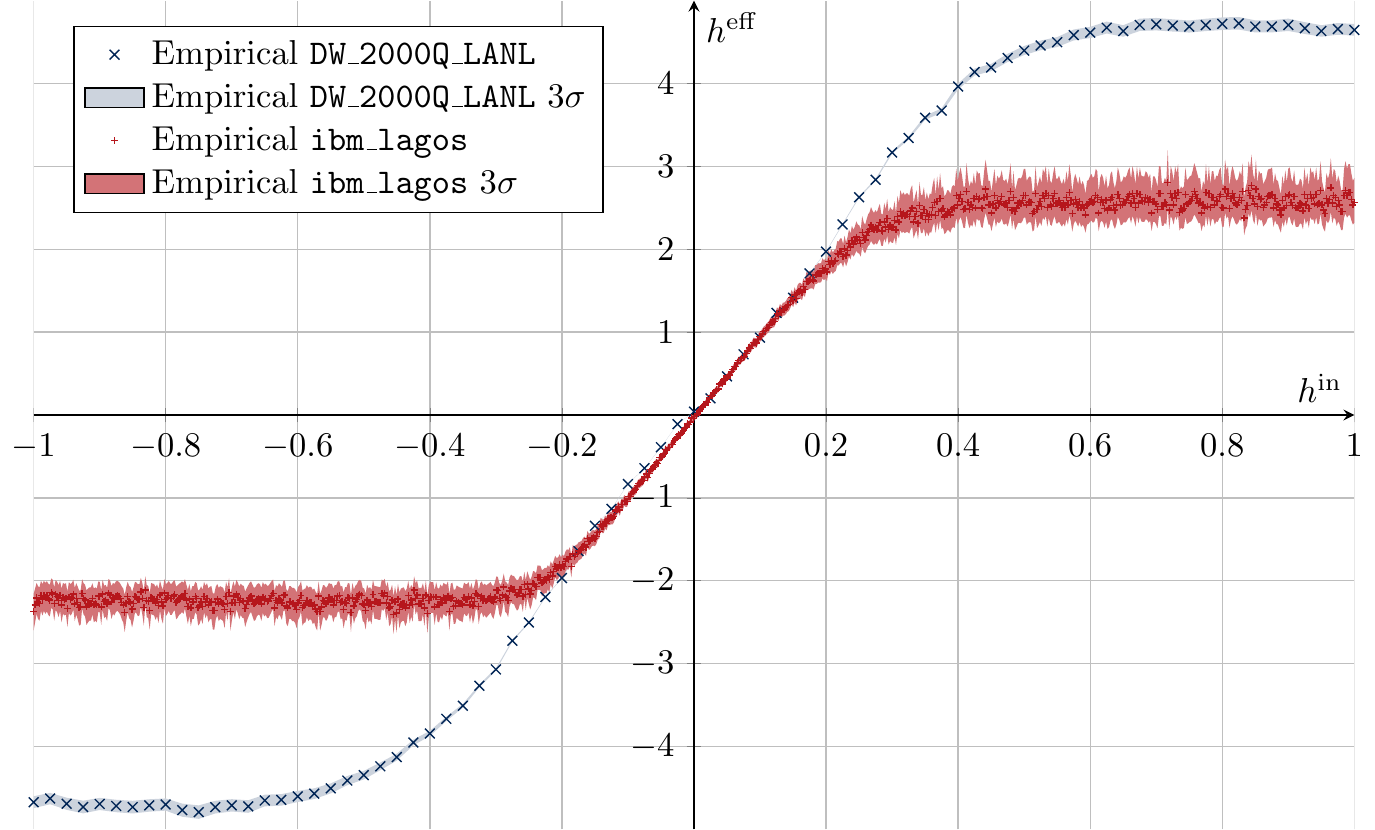}
        \caption{Data obtained after applying the protocol described in \cref{sec:theoretical-models} on two representative qubits, QC qubit ($6$) of \texttt{ibm\_lagos} and QA qubit ($305$) of \texttt{DW\_2000Q\_LANL}.}
        \label{fig:dwave-ibm}
    \end{subfigure}
    \hfill
    \begin{subfigure}[t]{0.49\textwidth}
        \centering
        \includegraphics[width=\textwidth]{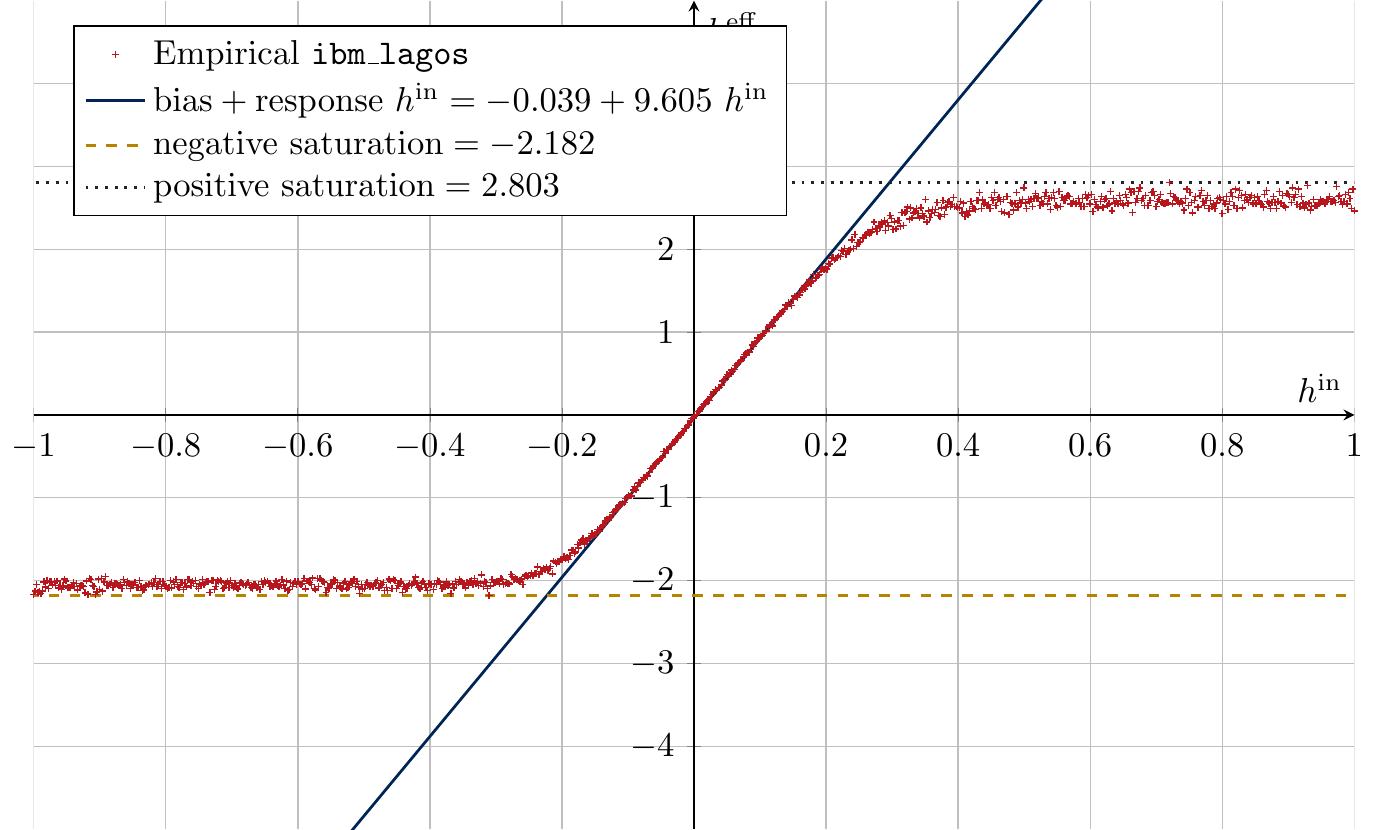}
        \caption{
        A visual representation of the $4$ features of interest to this work.
        The linear function $f$ is obtained by performing a linear regression on the \heff{} data in the interval \mbox{$\hin{} \in \left[ -0.1, 0.1 \right]$}. The $g$ (respectively $h$) function is obtained by computing the minimum and maximum \heff{} values over the whole range \mbox{$\hin{} \in \left[ -1, 1 \right]$}.}
        \label{fig:ibm-features}
    \end{subfigure}
    \caption{Comparison of the response values recovered from the Q-RBPN protocol on IBM-Q QC and D-Wave QA hardware. 900 and 81 evenly spaced $\hin{}$ values are shown for QC and QA respectively.}
    \label{fig:single-spin-experiment-with-features}
\end{figure}

To make this single-qubit evaluation procedure concrete, \cref{fig:dwave-ibm} provides an example of performing the complete protocol on representative qubits from the IBM-Q computer \texttt{ibm\_lagos} and the D-Wave 2000Q QA computer \texttt{DW\_2000Q\_LANL}.
In this illustration, a variety of input fields (i.e., $\hin{}$) ranging from $-1$ to $+1$ are executed on the hardware and the $h^{\text{eff}}$ values are computed.
Noting that the QA protocol of \texttt{DW\_2000Q\_LANL} yields $\beta \approx 10$ \cite{9465651,2012.08827}, in order to have comparable $\hin{}$/$\heff{}$ relationship between the two models of computation, we fix $\beta = 10$ in \cref{eq:theta-hin} of the QC model as well.

Recall from the discussion in \cref{sec:sq_models} that we expect a linear relationship between the values of $\hin{}$/$\heff{}$.
Reviewing the results presented in \cref{fig:dwave-ibm} largely confirms the expected linear relationship; however, one can observe notable deviations from the expected model including $\heff{}$ saturation effects and minor deviations in the linear response.
Two observations are of particular note: 
(1) the QA hardware realizes a maximal \heff{} magnitude around $5$, which is about twice as large as the maximum magnitude realized by the QC hardware ($\approx 2.5$); and 
(2) there is a notable asymmetry in the maximum magnitude of the negative ($\approx -2.0$) and positive ($\approx 2.5$) \heff{} values realized by the QC hardware.
The root source of these deviations and the differences between QC and QA platforms present interesting questions;
however, we will leave those investigations to future work and begin by proposing metrics that provide a coarse summary of a qubit's characteristics in this $\hin{}$/$\heff{}$ procedure.
As illustrated in \cref{fig:dwave-ibm}, the 4 metrics we propose are: {\em response / bias}, the slope and the offset extracted from fitting an affine model in the linear operating region of the qubit; and {\em positive / negative saturation}, the maximum and minimum observed $\heff{}$ values capturing the hardware's output saturation points.
To fit the affine model, a subset of the data is selected $\hin{} \in \left[ -0.1, 0.1 \right]$, where the response behaves linearly in both hardware platforms.
In general, this range may need to be revised based on the performance of a specific hardware device to yield and accurate fit of the data.

It is important to briefly remark on the data requirements for an accurate estimation of $h^{\textrm{eff}}$, especially for large $\hin{}$ values.
For each value of $\hin{}$, one collects $M$ samples to extract a conditional expectation $\widehat{\mathbb{E}}\left[\sigma \mid \hin{}\right]$, which corresponds to an empirical effective field $\heff{} = \tanh^{-1} \widehat{\mathbb{E}}\left[\sigma \mid \hin{}\right]$.
For a particular value of $M$, this estimator is subject to an accuracy limit caused by finite sampling.
For large values of $\vert h^{\textrm{eff}} \vert$, the probability of observing a qubit misaligned with the effective field decreases exponentially with the field's intensity; see \cref{eq:effective_outcome_prob}.
Therefore, if $h^{\textrm{eff}} = 5$ one only expects to see one misaligned spin configuration in every $22,000$ observations, requiring millions of samples to have a confident estimation of $h^{\textrm{eff}}$.
Hence, it is necessary to adjust these data collection requirements to be consistent with the hardware's performance.
This finite sampling accuracy challenge is addressed in this work by setting $M$ to a level that provides tight confidence intervals around the estimation of $h^{\textrm{eff}}$ for the particular QC and QA hardware that were considered; this resulted in $M = 8 \times 10^3$ and $M = 5 \times 10^6$ for each hardware platform respectively.
To that end, 99.7\% confidence intervals (i.e., 3 sigma) for the experimental data collected by this work are shown in \cref{fig:dwave-ibm}.

\vspace{-0.5cm}
\subsection{Limitations}

It worth briefly mentioning some of the limitations of this protocol and metrics proposed in this work.
It is widely known that current qubits suffer from notable amounts of state preparation and measurement errors (SPAM) \cite{Preskill2018quantumcomputingin} yielding spin flips on the order of 1 in 100 measurements \cite{ibmquantumwebsite}, which limits the \heff{} that can be observed on that hardware.
QA hardware is known to suffer from a wide range of {\it integrated control errors} (ICE) \cite{dwave_docs}, which include: background susceptibility; flux noise; Digital-to-Analog Conversion quantization; Input/Ouput system effects; and variable scale across qubits.
The differences in how these hardware errors impact qubit performance across these platforms presents a notable challenge for developing hardware agnostic performance metrics.
Although the metrics proposed in this work do not capture these specific known issues with qubit performance, they have the advantage of not requiring any assumptions about the particular computing model or underlying qubit implementation.
Developing approaches for capturing the signatures of a wider variety of qubit errors presents an opportunity for expanding the Q-RBPN protocol in future work.

\section{Full-Chip Parameter Comparisons}
\label{sec:full_chip}

The ultimate goal of this work is to develop scalable qubit performance metrics that can be used across both QC and QA computing platforms. 
To that end, we observe that the single-qubit protocol discussed in the previous section can be executed in parallel for every qubit in QC and QA hardware devices.
Consequently, we propose the Qubit Response, Bias, Positive saturation, Negative saturation (Q-RBPN) protocol that runs the $\hin{}$/$\heff{}$ analysis in parallel characterizes the distribution of response, bias, positive/negative saturation, across an entire quantum computing hardware device.
This protocol enables the comparison of a large numbers of qubits spanning a range of quantum hardware to highlight the consistency of qubits across systems, not just individual qubit comparisons.

To explore the potential of the Q-RBPN protocol, this work analyzes 2300 qubits spanning 18 QC platforms in IBM's Q-Hub (311 qubits in total) and a D-Wave 2000Q Quantum Annealer located at Los Alamos National Laboratory (2031 qubits), known as \texttt{DW\_2000Q\_LANL}.  A summary of all of the quantum computers considered in this work and their Q-RBPN results is available in \cref{tab:ibmq_chip_data}. 

On the IBM-Q hardware, the following data collection settings were used: the number of shots performed is $8192$, and the $1$-qubit circuits have all been compiled with a custom transpiler pass that guarantees that the result of the compilation will always include $5$ hardware-native gates, which covers the most general case for a single qubit. 
This custom transpilation pass is needed to avoid certain special-cases where the $5$ gate decomposition can be simplified to $3$ gates or even a single $R_z$ gate, which would distort the results obtained by the protocol as executing fewer gates results in lower error rates.
See \cref{sec:gate-consistency} for additional details around the $5$ gate decomposition used in this work.

On the D-Wave hardware the following  data collection settings were used, {\it flux drift compensation} is disabled, which prevents automatic corrections to input fields based on a calibration procedure that is run a few times each hour; the {\it num reads} is set to 10000, specifying the number of identical executions performed for a single programming cycle of the chip; and the {\it annealing time} is set to 1 $\mu s$. For each $h^{\text{in}}$ input value, the $h^{\text{eff}}$ is estimated with $5 \times 10^6$ identical executions to ensure high accuracy of the $h^{\text{eff}}$ estimation.
The following sections discuss the results of running this Q-RBPN protocol on all 2300 qubits considered revealing some insights into the variations between these two quantum computing platforms.

\subsection{Parameter Distributions}

\begin{figure}[t]
    \centering
    \begin{subfigure}[b]{0.45\textwidth}
        \centering
        \includegraphics[width=\textwidth]{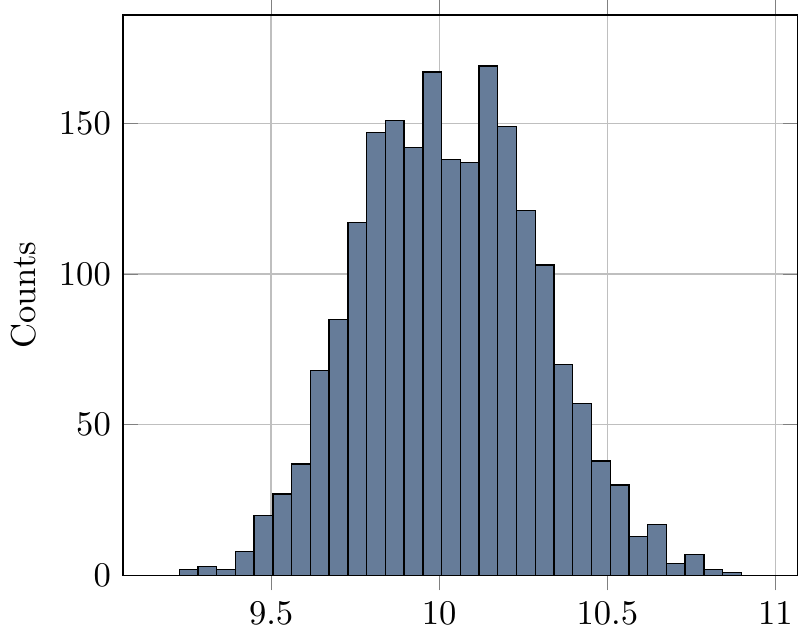}
        \caption{Histogram of fitted response values on D-Wave \mbox{($n=2031$)}.}
        \label{fig:dwave_slopes}
    \end{subfigure}
    \hfill
    \begin{subfigure}[b]{0.45\textwidth}
        \centering
        \includegraphics[width=\textwidth]{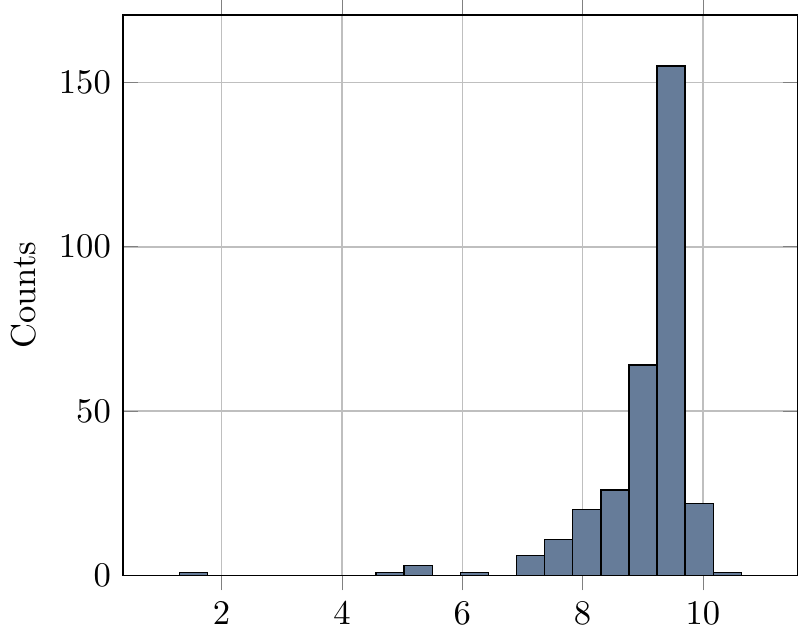}
        \caption{Histogram of fitted response values on IBM-Q \mbox{($n=311$)}.}
        \label{fig:ibmq_slopes}
    \end{subfigure}
    \caption{Comparison of the response values recovered from the Q-RBPN protocol on D-Wave and IBM-Q hardware.}
    \label{fig:hist-comparisons-slope}
\end{figure}

\paragraph*{Response:}
The response parameter recovered by the Q-RBPN protocol is arguably the most important value as it reflects the hardware accuracy in responding to a given input value and highlights the $h^{\text{in}}$ parameter range where the hardware will exhibit the best performance.
\Cref{fig:hist-comparisons-slope} provides a summary of the response values recovered from the QA (top) and QC (bottom) hardware platforms.
Both platforms provide a range of values for the response with mean and standard deviations of 10.03 $\pm$ 0.26 (QA) and 9.04 $\pm$ 0.88 (QC).
These specific mean values are not of particular note as they can be adjusted in both platforms by changing how the quantum program is executed, see \cref{eq:theta-hin} and \cite{9465651}.
However, the variance value is interesting because it reflects the variability of the response value across a cohort of qubits that ideally would have identical response functions.
These recovered variance values suggest that the QC qubits exhibit around twice the variability in the response than the QA qubits, on average.
Given that averages over multiple QC chips may be misleading, we suggest reviewing the chip-by-chip response variance values in \cref{tab:ibmq_chip_data} for a more nuanced comparison where some QC hardware is able to match the QA response variability and some have notably higher values.

\begin{figure}[t]
    \centering
    \begin{subfigure}[t]{0.45\textwidth}
        \centering
        \includegraphics[width=\textwidth]{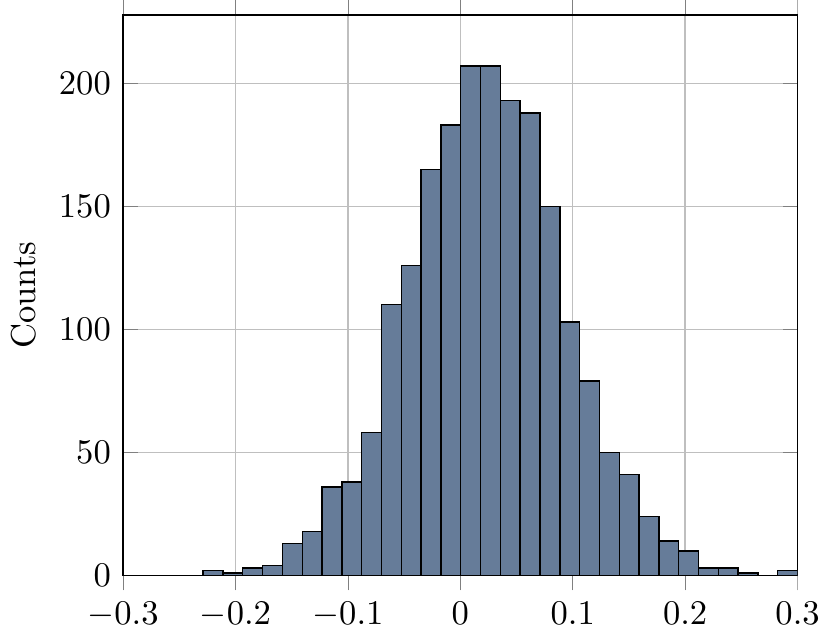}
        \caption{Histogram of the computed biases for all the qubits on the D-Wave chip ($n=2031$).}
        \label{fig:dwave_offsets}
    \end{subfigure}
    \hfill
    \begin{subfigure}[t]{0.45\textwidth}
        \centering
        \includegraphics[width=\textwidth]{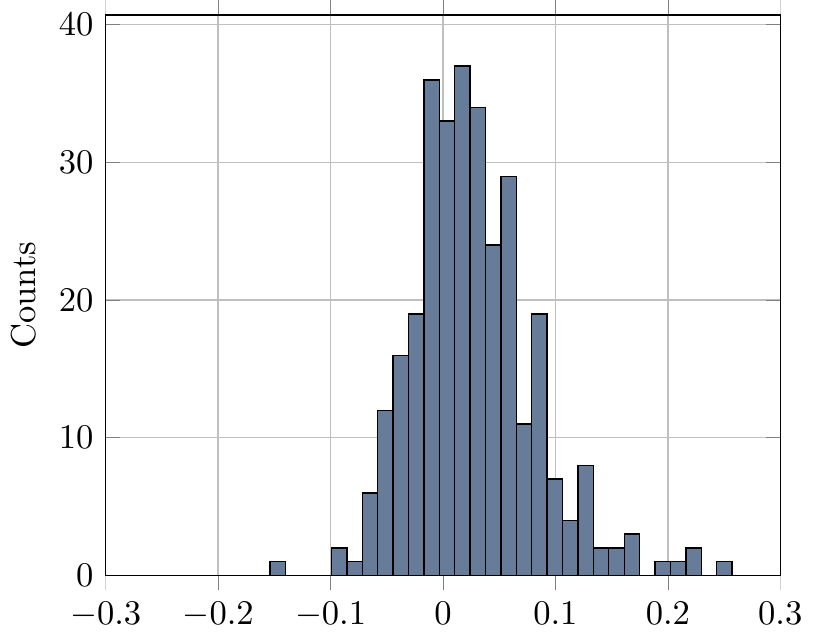}
        
        \caption{Histogram of the computed biases for all the qubits on all the IBM-Q chips ($n=311$).}
        \label{fig:ibmq_offsets}
    \end{subfigure}
    \caption{Comparison of the bias distribution between the quantum-annealing model of computation with D-Wave chip and the gate-based model of computation with IBM-Q hardware.  For consistency both plot's axis range of -0.3 to 0.3, which excludes three outliers from the D-Wave dataset.}
    \label{fig:hist-comparisons-offset}
\end{figure}

\paragraph*{Bias:}
The bias parameter recovered by the Q-RBPN protocol is arguably the least important value as can easily be corrected by prepossessing the input data.
However, its magnitude and variability do provide a measure of the calibration accuracy of the qubits.
\Cref{fig:hist-comparisons-offset} provides a summary of the bias values recovered from the QA (top) and QC (bottom) hardware platforms.
Both platforms provide a fairly narrow range of bias values with mean and standard deviations of 0.02 $\pm$ 0.07 (QA) and 0.03 $\pm$ 0.06 (QC).
Although the two histograms appear to be distinct, computation of the mean and standard deviation reveals that their performance is remarkably similar.
This level of similarity is notable given the desperate implementations of these two types of quantum devices.
One possible explanation is that both platforms are built on super conducting loops operating a milli-Kelvin temperatures; possibly this underlying technology is the limiting the calibration accuracy of these qubits.
Executing the Q-RBPN protocol on a quantum hardware platform that uses a different foundational technology (e.g., trapped ions) could provide additional insights into this hypothesis and is left for future work. 

\begin{figure*}[t]
    \centering
    \begin{subfigure}[b]{\textwidth}
        \centering
        \includegraphics[width=\textwidth]{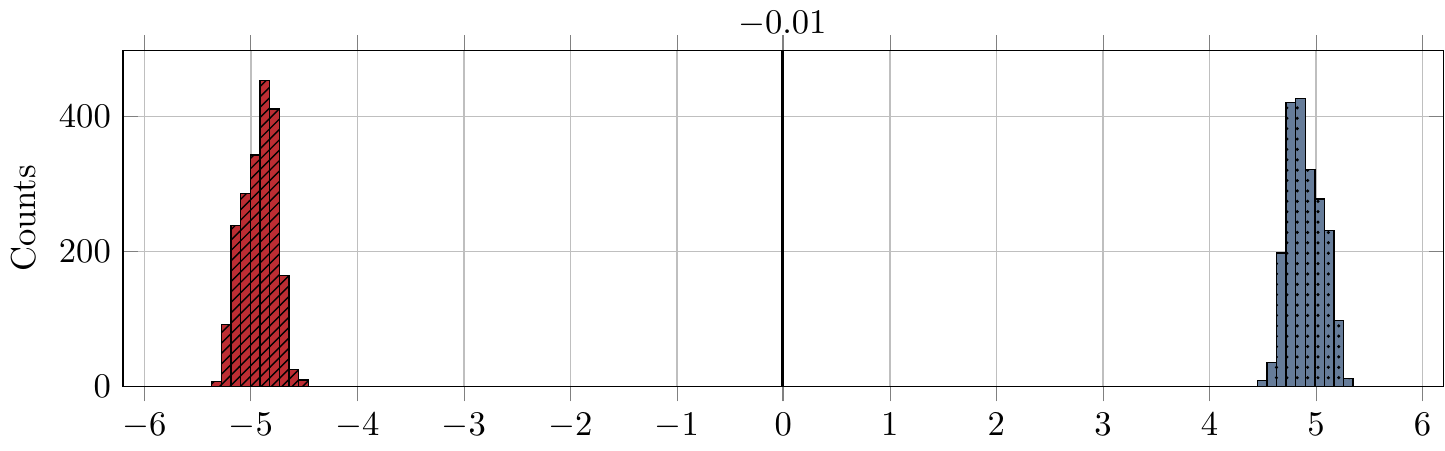}
        \caption{Histogram of the negative and positive saturation values on D-Wave ($n=2031$).}
        \label{fig:dwave_mins}
    \end{subfigure}
    \newline
    \begin{subfigure}[b]{\textwidth}
        \centering
        \includegraphics[width=\textwidth]{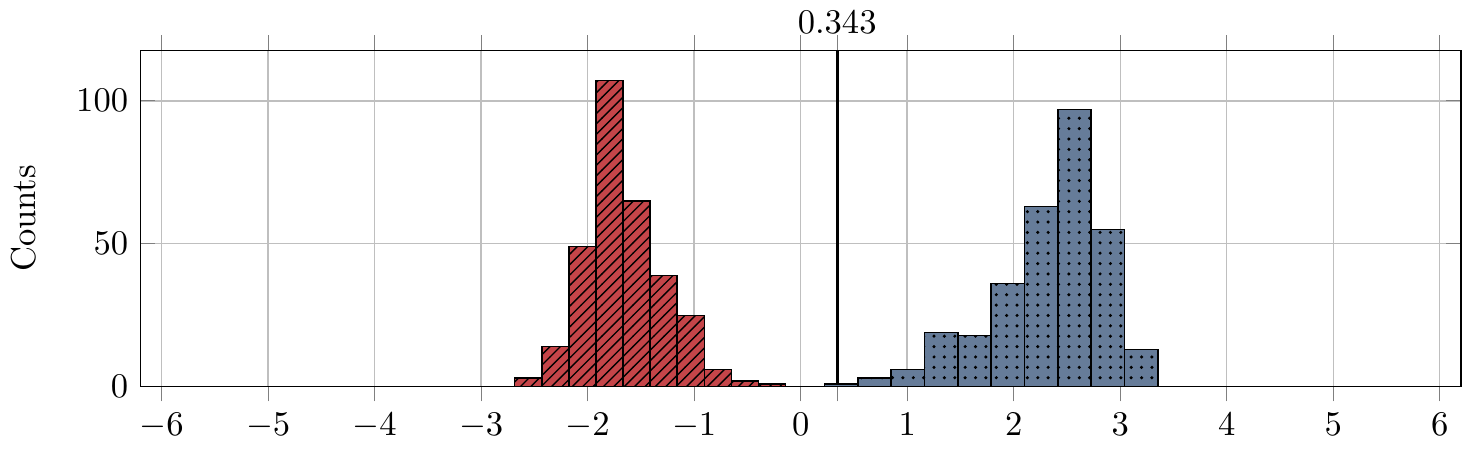}
        \caption{Histogram of the negative and positive saturation values on IBM-Q ($n=311$).}
        \label{fig:ibmq_mins}
    \end{subfigure}
    \caption{Comparison of the distributions of negative and positive saturation for D-Wave hardware (above) and IBM-Q hardware (below). The vertical black line on each plot corresponds to the mean over all the collected data for the chip considered. The D-Wave mean of $-0.01$ indicates that the negative/positive saturation values are distributed in a nearly symmetric fashion. The $0.343$ mean value for IBM-Q hardware indicates that the negative/positive saturation distributions are biased tower positive values.}
    \label{fig:hist-comparisons-min-max}
\end{figure*}

\paragraph*{Positive / Negative Saturation:}
The saturation values recovered by the Q-RBPN protocol provide useful information on the range of input parameters that can be realized in practice.
For example, notice in \cref{fig:dwave-ibm} that there is no effective difference in $\heff{}$ for the range of $|\hin{}| \in [0.6, 1.0]$, suggesting a programmable range of $\hin{} \in [-0.4, 0.4]$ would be more suitable if the user would like to the hardware respond consistently to different $\hin{}$ values.

\Cref{fig:hist-comparisons-min-max} provides a summary of the saturation values recovered from the QA (top) and QC (bottom) hardware platforms.
This parameter highlights the biggest differences in both of these hardware platforms.
A first observation is that QA hardware realizes notably larger saturation values ($\approx 5.0$) than the QC hardware ($\approx 2.5$).
This result suggests a larger programmable range for the QA hardware as compared to the QC hardware.
Note that using the protocol presented in \cref{sec:sq_models} to realize large values of $\heff{}$ requires the QC hardware to resolve very fine angles (i.e., to distinguish values of $\heff{} \geq 5$ requires an angle resolution below 0.01 radians), presenting a significant test of the hardware's control system.
A second observation is that the variation of saturation points of the QA hardware ($\approx 0.15$) tend to be more consistent that the QC hardware ($\approx 0.30$).
Finally, these results highlight that the asymmetry of the saturation values in the QC hardware that are shown in \cref{fig:dwave-ibm} are a fairly consistent feature of the IBM-Q hardware.
We suspect this asymmetry is related to operational challenges that encourage the hardware to prefer one state over another; however, a comprehensive test of this hypothesis is left for future work.
For application developers, knowledge of this QC hardware asymmetry is useful as it can inspired different treatment of negative and positive \hin{} values. 
Overall, these saturation metrics provide a clear distinction between the qubits in the QA and QC hardware considered here and highlights one feature for possible improvement in QC hardware.

\newcolumntype{M}[1]{>{\raggedleft\arraybackslash}m{#1}}
\begin{table*}[]
    \centering
    \begin{tabular}{|l|r||M{2cm}|M{2cm}|M{2cm}|M{2cm}|}
      \hline
      {\bf Chip Name} & {\bf Qubits} & {\bf Response} & {\bf Bias} & {\bf Negative S.} & {\bf Positive S.} \\
      \hline
      \hline
      \texttt{DW\_2000Q\_LANL}& $2031$ & $10.03 \pm 0.26$& $0.02 \pm 0.07$& $-4.92 \pm 0.15$& $4.90 \pm 0.16$\\
      \hline
      all ibmq backends & $311$ & $9.04 \pm 0.88$& $0.03 \pm 0.06$& $-1.65 \pm 0.37$& $2.33 \pm 0.53$\\
      \hline
      \hline
      \texttt{ibmq\_brooklyn} & 65& $9.21 \pm 0.57$& $0.02 \pm 0.05$& $-1.68 \pm 0.29$& $2.38 \pm 0.32$\\
      \hline
      \texttt{ibmq\_manhattan} & 65& $8.95 \pm 0.86$& $0.03 \pm 0.06$& $-1.60 \pm 0.40$& $2.24 \pm 0.56$\\
      \hline
      \texttt{ibmq\_montreal} & 27& $9.23 \pm 0.57$& $0.02 \pm 0.06$& $-1.80 \pm 0.33$& $2.47 \pm 0.48$\\
      \hline
      \texttt{ibmq\_mumbai} & 27& $8.96 \pm 0.60$& $0.01 \pm 0.03$& $-1.61 \pm 0.25$& $2.06 \pm 0.47$\\
      \hline
      \texttt{ibmq\_sydney} & 27& $8.73 \pm 1.58$& $0.04 \pm 0.05$& $-1.54 \pm 0.36$& $2.25 \pm 0.68$\\
      \hline
      \texttt{ibmq\_toronto} & 27& $8.80 \pm 1.21$& $0.04 \pm 0.05$& $-1.67 \pm 0.48$& $2.35 \pm 0.72$\\
      \hline
      \texttt{ibmq\_guadalupe} & 16& $9.42 \pm 0.16$& $0.03 \pm 0.05$& $-1.75 \pm 0.16$& $2.75 \pm 0.29$\\
      \hline
      \texttt{ibmq\_casablanca} & 7& $9.38 \pm 0.28$& $0.04 \pm 0.03$& $-1.73 \pm 0.23$& $2.71 \pm 0.22$\\
      \hline
      \texttt{ibmq\_jakarta} & 7& $9.19 \pm 0.31$& $0.03 \pm 0.06$& $-1.65 \pm 0.26$& $2.32 \pm 0.27$\\
      \hline
      \texttt{ibm\_lagos} & 7& $9.79 \pm 0.41$& $-0.02 \pm 0.03$& $-2.32 \pm 0.36$& $2.50 \pm 0.23$\\
      \hline
      \texttt{ibmq\_belem} & 5& $9.09 \pm 0.35$& $0.10 \pm 0.06$& $-1.43 \pm 0.22$& $2.39 \pm 0.31$\\
      \hline
      \texttt{ibmq\_bogota} & 5& $9.08 \pm 0.68$& $0.05 \pm 0.04$& $-1.58 \pm 0.39$& $2.53 \pm 0.33$\\
      \hline
      \texttt{ibmq\_lima} & 5& $9.08 \pm 0.51$& $0.10 \pm 0.08$& $-1.45 \pm 0.38$& $2.78 \pm 0.18$\\
      \hline
      \texttt{ibmq\_manila} & 5& $9.38 \pm 0.06$& $0.00 \pm 0.02$& $-1.68 \pm 0.06$& $2.35 \pm 0.29$\\
      \hline
      \texttt{ibmq\_quito} & 5& $9.05 \pm 0.46$& $0.08 \pm 0.03$& $-1.38 \pm 0.22$& $2.48 \pm 0.44$\\
      \hline
      \texttt{ibmq\_santiago} & 5& $8.95 \pm 0.50$& $0.05 \pm 0.02$& $-1.57 \pm 0.33$& $2.15 \pm 0.50$\\
      \hline
      \texttt{ibmqx2} & 5& $7.31 \pm 1.03$& $-0.04 \pm 0.08$& $-1.16 \pm 0.24$& $1.30 \pm 0.44$\\
      \hline
      \texttt{ibmq\_armonk} & 1& $8.23 \pm 0.00$& $0.01 \pm 0.00$& $-1.28 \pm 0.00$& $1.41 \pm 0.00$\\
      \hline
    \end{tabular}
    \caption{A summary of the Q-RBPN metrics broken down by specific hardware devices including one D-Wave quantum annealing hardware and eighteen IBM-Q quantum computing hardware. The statistics of each of the Q-RBPN metrics are presented in the format \texttt{[mean]} $\pm$ \texttt{[standard deviation]}.}
    \label{tab:ibmq_chip_data}
\end{table*}

\section{Discussion \& Conclusion}
\label{sec:conclusion}

In the era of NISQ devices, QCVV plays a valuable role in measuring and tracking the fidelity of quantum hardware platforms.
However, QCVV protocols that can measure and compare the performance of both QC and QA hardware have been limited.
In this work, we proposed the Q-RBPN protocol to fill this gap and provide a simple and scalable single-qubit performance comparison across different quantum hardware platforms.
We hope that the four metrics proposed in this work -- response, bias, positive/negative saturation -- will provide valuable tools for tracking the progress of a wide range of quantum hardware platforms over time, irrespective of the computational model under consideration.

In addition to tracking qubit performance, the metrics proposed in this work also have interesting implications for specific applications of quantum computing.
In the context of quantum accelerated Boltzmann sampling, the saturation analysis in this work provides insights into what types of distribution parameters can be accurately represented on a given hardware device.
If we consider calibration of variational quantum algorithms with hardware in the loop, these metrics provide bounds on the parameter values where the hardware has a good response to the input functions.
In the context of quantum accelerated optimization, the programmer often has freedom to rescale the problem parameters.  The saturation values provided by Q-RBPN indicate the optimal problem scaling on a given hardware platform, which can increase the accuracy of cross-platform optimization performance comparisons.

Finally, this work has revealed some key directions for future investigation.
The first direction is to execute the Q-RBPN protocol on a wider range of quantum hardware platforms \cite{Ladd2010}.
Ion trap based QC \cite{,Debnath2016} and emerging QA \cite{PRXQuantum.01.020314} hardware platforms are natural next steps to understand how the proposed metrics vary across additional hardware realizations.
A second direction is to better understand what are the underlying mechanisms that result in non-ideal behavior of qubits in the Q-RBPN protocol.
Such insights could improve the interpretability of the Q-RBPN metrics, introduce additional measures of qubit performance and potentially point to fundamental limits on the performance of the underlying hardware.

\section{Acknowledgements}

This work was partly supported by the U.S. DOE through a quantum computing program sponsored by the Los Alamos National Laboratory (LANL) Information Science \& Technology Institute and the Laboratory Directed Research and Development (LDRD) program of LANL under project numbers 20210116DR and 20210114ER. This research was also partly supported by the U.S. Department of Energy (DOE), Office of Science, Office of Advanced Scientific Computing Research, under the Accelerated Research in Quantum Computing (ARQC) program.
This research used quantum computing resources provided by the LANL Institutional Computing Program, which is supported by the U.S. Department of Energy National Nuclear Security Administration under Contract No. 89233218CNA000001.
Adrien Suau also thanks Total Energies for their general support of this work.

\bibliography{main}

\vspace{0.5cm}
LA-UR-21-28409

\clearpage
\appendix

\section{Alternative Quantum Gate Programs}
\label{sec:alternative-quantum-gate-program}

\begin{figure}[t]
    \centering
    \includegraphics{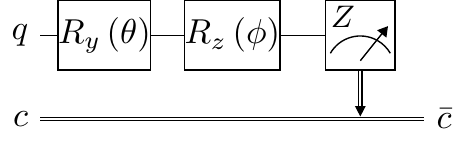}
    \caption{A more general $1$-qubit quantum program used to conduct the Q-RBPN protocol. The rotation angle $\phi$ is a free input parameter and the rotation angle $\theta$ is obtained from specified \hin{} values via \cref{eq:theta-hin}.}
    \label{fig:one-qubit-qc-phi-theta}
\end{figure}

The quantum circuit presented in \cref{fig:one-qubit-qc-theta} is suitable for implementing the Q-RBPN protocol developed in this work. 
However, this is not the only program that could have been used in QC context. 
In particular, this $1$-parameter circuit is only able to explore states that are in the X-Z plane of the Bloch sphere.
It is reasonable to postulate that more sophisticated QC programs may yield different results in the Q-RBPN protocol. 
One option for considering a more general QC program is to perform the same $1$-parameter experiment but in another plane, for example, in the Y-Z plane. 
This process is accomplished by adding an $R_z\left( \phi \right)$ rotation that will rotate the plane in which the experiment is performed. For the Y-Z plane, $\phi=\frac{\pi}{2}$. The resulting quantum circuit is depicted in \cref{fig:one-qubit-qc-phi-theta}.

\begin{figure}[t]
    \centering
    \includegraphics[width=0.49\textwidth]{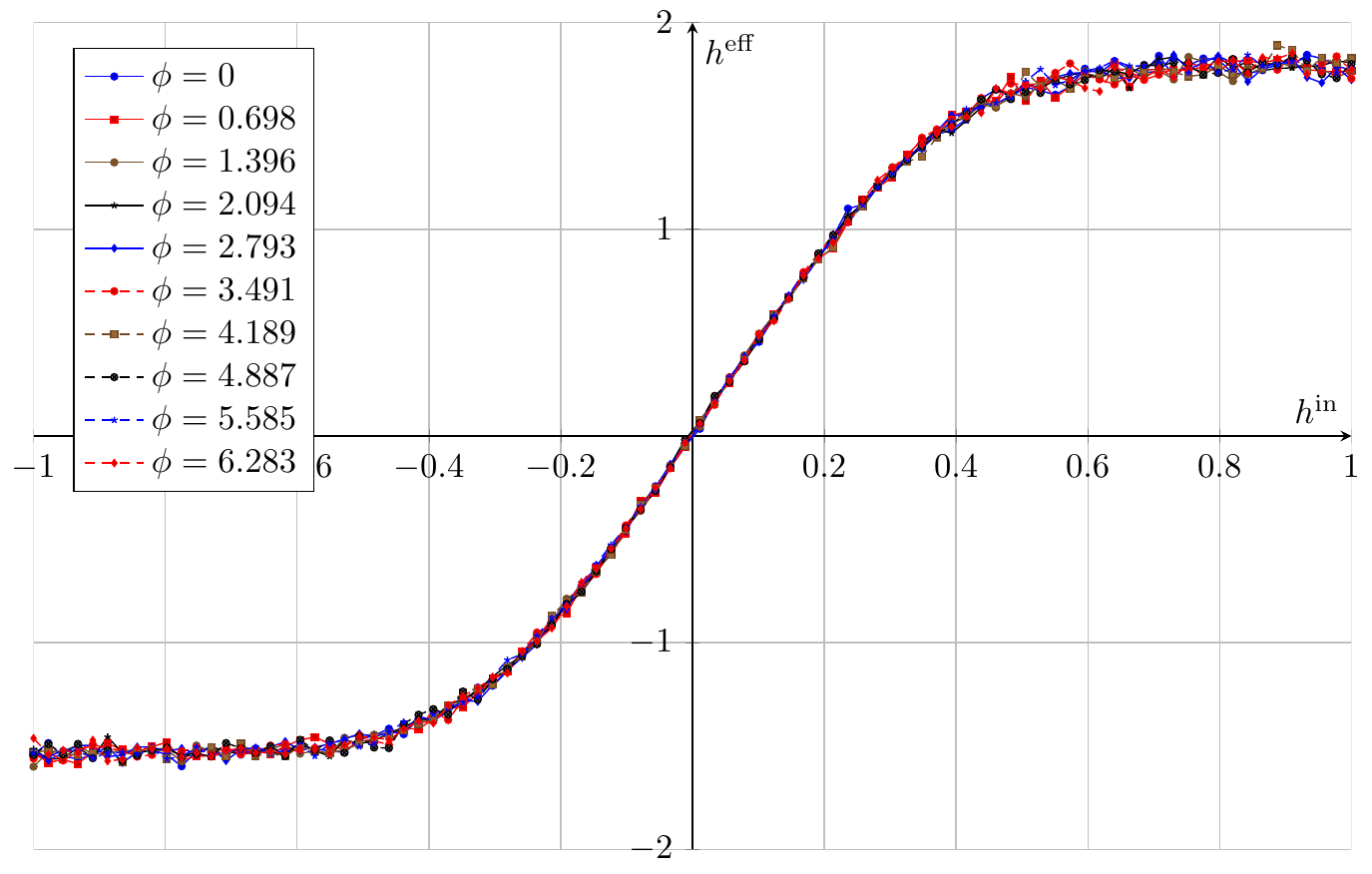}
    \caption{Result with multiple different values for $\phi$ zoomed. Data gathered on qubit $0$ of \texttt{ibm\_lagos}. Plot obtained with a value of $\beta = 10$.}
    \label{fig:sspin_multiple_phis_zoomed}
\end{figure}

To explore the potential of performing the experiments in a different plane, we repeat the Q-RBPN protocol for a single qubit with 10 evenly spaced values of $\phi \in [0, 2\pi]$ and the results are presented in \cref{fig:sspin_multiple_phis_zoomed}.
These results are what is expected theoretically as an $R_z\left( \phi \right)$ gate applied just before the qubit measurement has no effect on the expected outcome of the measurement.
In particular, the $R_x(\theta)$ rotation is simply an $R_y\left( \theta \right)$ rotation followed by a $R_z\left( \frac{\pi}{2} \right)$ rotation.
Based on the consistency of these results we conclude that the simpler quantum circuit presented in \cref{fig:one-qubit-qc-theta} is suitable for implementing the Q-RBPN protocol developed in this work.

\begin{figure}[h!]
    \vspace{0.40cm}
    \centering
    \includegraphics{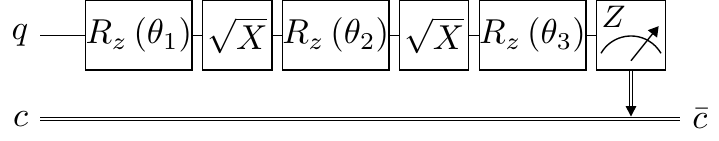}
    \caption{The $1$-qubit quantum program in the IBM-Q native gate set that was used in this work to perform any single qubit operation with a standardized program size.}
    \label{fig:one-qubit-qc}
\end{figure}

\section{Quantum Gate Program Consistency}
\label{sec:gate-consistency}

Given the NISQ nature of available QC quantum hardware, the specific details of the quantum programs can have a notable impact on performance~\cite{cross2017open, cincio2020machine}.
When performing benchmarking and extracting metrics from several different quantum circuits, consistency is paramount; i.e., one does not want a few quantum circuits being treated differently than the others.
In particular, it is important in the Q-RBPN protocol that all of the quantum circuits used to perform an experiment have the same number and type of quantum gates.
This configuration is important because each gate introduces some amount of error into the computation \cite{ibmquantumwebsite}, and, hence, executing smaller circuits changes overall error model.
Running identical circuits reduces the potential biases that might emerge when different equivalent circuits are executed.

As this work is only concered with benchmarking individual qubits, we observed that there exists a 5 gate program (using IBM-Q native gates) that can implement any $1$-qubit operation as illustrated in \cref{fig:one-qubit-qc}.
This 5 gate program tends to also be the smallest program that can implement a given single-qubit operation; however, some care needs to be taken with special-cases that arise; for example, when the $1$-qubit quantum gate can be compiled to a single $R_z\left( \theta \right)$ gate.
In such cases, for consistency, we want the quantum circuit to still contain the $5$ gates from \cref{fig:one-qubit-qc}, even though $1$ gate would be enough. This result is achieved by implementing a specific transpilation pass for IBM's qiskit transpiler, to ensure that all $5$-gates are generated for any desired single-qubit operation.

\end{document}